%% file: paper.tex
\documentclass[10pt,conference]{IEEEtran}
\usepackage{cite}
\usepackage{amsmath,amssymb,amsfonts}
\usepackage{algorithmic}
\usepackage{graphicx}
\usepackage{textcomp}
\usepackage{color,soul}
\def\BibTeX{{\rm B\kern-.05em{\sc i\kern-.025em b}\kern-.08em
    T\kern-.1667em\lower.7ex\hbox{E}\kern-.125emX}}

\newcommand{\australia}{Australia}
\newcommand{\malaysia}{Malaysia}
\newcommand{\monash}{Monash University}
\newcommand{\checklist}{checklist}

\newcommand\checklistItem[1]{question#1}

\newcommand\rqOne[1]{To what degree can the students’ checklists be used to identify defects during code review?}

\newcommand\rqTwo[1]{What are the common mistakes in students' checklists?}

\newcommand{\ea}{\textit{et al.}}

\begin{document}
\title{Assessing the Students' Understanding and their Mistakes in Code Review Checklists \\ \Large{--An Experience Report of 1,791 Code Review Checklist Questions from 394 Students--}}


\author{\IEEEauthorblockN{Chun Yong Chong}
\IEEEauthorblockA{Monash University, Malaysia. \\
chong.chunyong@monash.edu
}
\and
\IEEEauthorblockN{Patanamon Thongtanunam}
\IEEEauthorblockA{The University of Melbourne, Australia. \\
patanamon.t@unimelb.edu.au
}
\and
\IEEEauthorblockN{Chakkrit Tantithamthavorn}
\IEEEauthorblockA{Monash University, Australia. \\
chakkrit@monash.edu
}
}



\maketitle

\begin{abstract}
Code review is a widely-used practice in software development companies to identify defects.
Hence, code review has been included in many software engineering curricula at universities worldwide.
However, teaching code review is still a challenging task because the code review effectiveness depends on the code reading and analytical skills of a reviewer.
While several studies have investigated the code reading techniques that students should use to find defects during code review, little has focused on a learning activity that involves analytical skills.
Indeed, developing a code review checklist should stimulate students to develop their analytical skills to anticipate potential issues (i.e., software defects). 
Yet, it is unclear whether students can anticipate potential issues given their limited experience in software development (programming, testing, etc.).
We perform a qualitative analysis to investigate whether students are capable of creating code review checklists, and if the checklists can be used to guide reviewers to find defects. 
In addition, we identify common mistakes that students make when developing a code review checklist.
Our results show that while there are some misconceptions among students about the purpose of code review, students are able to anticipate potential defects and create a relatively good code review checklist.
Hence, our results lead us to conclude that developing a code review checklist can be a part of the learning activities for code review in order to scaffold students' skills. 
\end{abstract}


\maketitle

\section{Introduction}

Code review has been widely known to be an effective software quality assurance technique to detect defects through a manual examination of source code written by different developers, other than the code authors themselves~\cite{fagan2002design,thongtanunam2015investigating,bacchelli2013expectations,thongtanunam2015should}. 
While the goals of automated software testing are much more focused on the functional correctness of software systems, the goals of code reviews include broader defect types such as the reliability of software architecture, the maintainability of source code, and the volatility of code changes \cite{Aurum2002,fagan2002design}.
Shull~\ea~argued that code reviews can uncover more than half of software defects which could be extra-deterministic behaviour and such defects cannot be detected by software testing~\cite{Shull2002}. 
Several studies found that code reviews have an impact on software quality~\cite{thongtanunam2016revisiting,thongtanunam2020tse}.
Moreover, prior work also showed the quality of other software artifacts such as requirements, design documents, as well as test cases~\cite{dunsmore2003development, Gilb1993, Fagan1976}.
Code reviews have been widely used in both open source and proprietary organisations (e.g., Google, Microsoft, Facebook, and OpenStack)~\cite{sadowski2018modern,Rigby,Shimagaki2016}.
In addition, code review has also been used as an assessment method for recruiting software engineers in a software company.\footnote{https://vampwillow.wordpress.com/2015/03/05/hacking-thoughtworks-recruitment-revisited-graduate-code-reviews/}

\begin{figure*}[t]
\centering
\includegraphics[width=.7\textwidth]{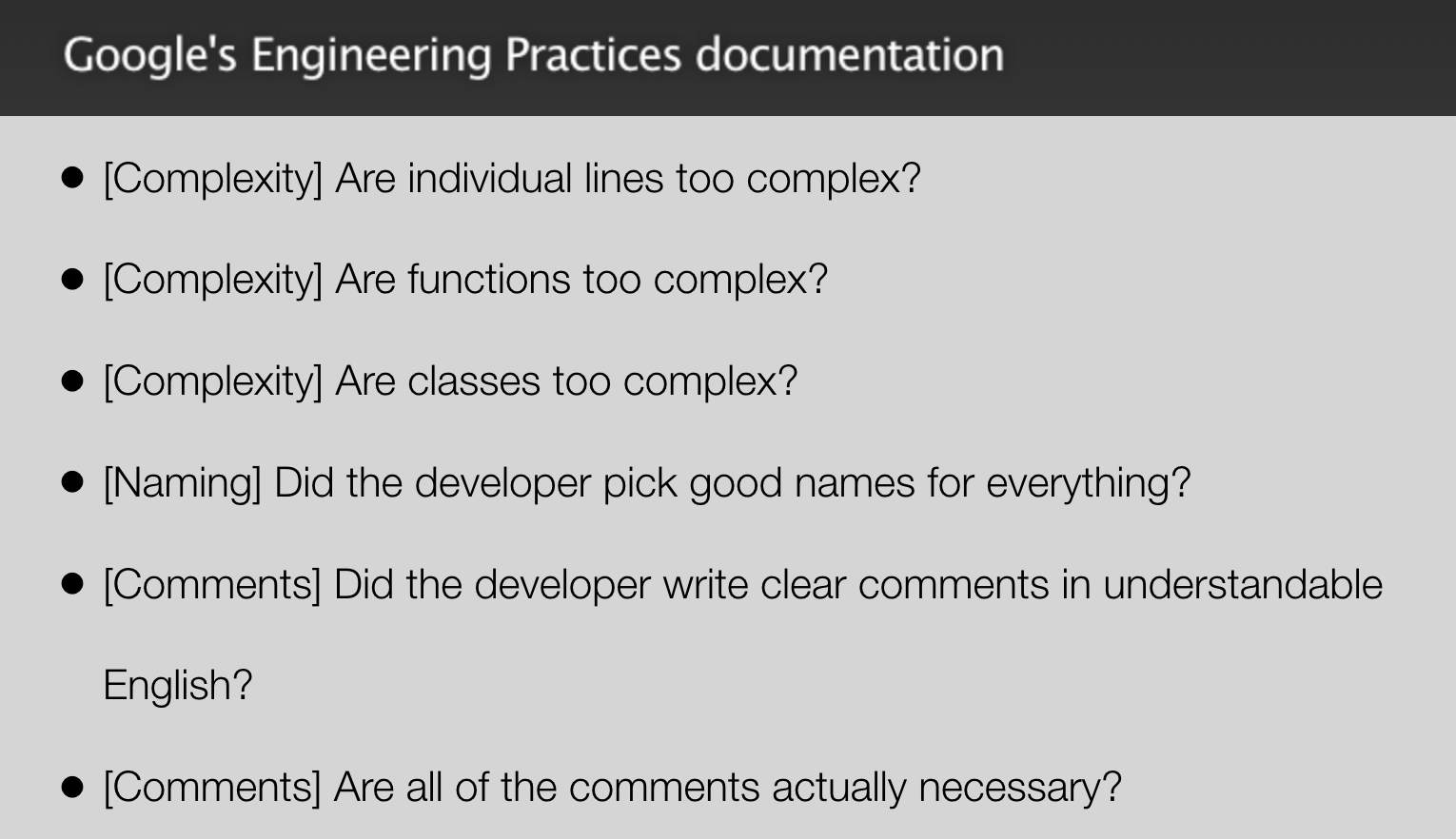}
\caption{An example of code review checklists from Google's Engineering Practices documentation}
\label{fig:fagan_sample}
\end{figure*}

Since code review has become an important and mandatory Software Engineering (SE) practice in the software industry, it is imperative for SE students to be equipped with code review skills prior to employment. 
However, when teaching code reviews to students, one of the main challenges is modelling students' critical software analysis and reading skills, which are the important skillsets for conducting code reviews. 
To perform effective code reviews, students should (1) anticipate the common issues or defects in software development, (2) prioritize the importance and severity of the anticipated issues, and (3) analyse software artefacts to identify potential issues (i.e., software defects)~\cite{Fagan1976}.
Yet, the common teaching and learning approaches are focused on the last step, i.e., analysing software artefacts by asking students to examine a given piece of source code that is embedded with defects~\cite{mantyla2008types, wang2012assessment}.
However, such a code review activity requires a lot of technical knowledge and support to help students to achieve the targeted goal (i.e. maximise the number of defects found during a code review), which can be a cognitive overload for students especially those with less experience in programming~\cite{mcmeekin2009evaluating}.

Prior studies show that checklist-based code review is an effective method of learning code reviews for inexperienced students, since the checklist can guide students to focus on the important issues or defects when performing code reviews~\cite{dunsmore2003development,rong2012effect}.
Checklist-based code review is considered to be more systematic than \emph{ad-hoc} reviewing (i.e., students identify defects from given software artefacts without checklists and proper instructions)~\cite{Aurum2002}. Typically, a checklist is prepared for a specific type of software artifact (code, UML, UI mockup, etc.), where the checklist may contain multiple questions related to the artifact to be examined. Hence, "a question" or "checklist questions" to refer the item(s) in the checklist.
Such a systematic code review method is more suitable for less experienced students~\cite{mcmeekin2009evaluating}.
Checklist-based code review is one of the structured code review techniques that was formalised by Fagan~\cite{fagan2002design,Aurum2002}.
A \emph{checklist} is a set of questions related to the potential issues or defects that might occur in code reviews.
Figure~\ref{fig:fagan_sample} shows an example of code review checklists from Google's Engineering Practices documentation.\footnote{https://google.github.io/eng-practices/review/reviewer/looking-for.html}


Although checklist-based code review is an effective method to train students' reading skills~\cite{rong2012effect}, providing \checklist{s} to students may not allow students to develop their analytical skills to anticipate potential issues or defects since they might directly use the provided checklist without understanding and appreciating the key motivations behind checklist-based code review.
In addition, the current code review practice in the industry setting generally does not provide code checklists. Instead, reviewers should depend on their own experience and domain knowledge to anticipate issues (i.e., having a checklist in mind) when performing code reviews~\cite{sadowski2018modern}. 
Therefore, having students to develop their own \checklist{s} should stimulate students to develop the critical and analytical skills to be able to anticipate potential issues or defects in software artefacts that the students will review.
Yet, it is unclear whether students can anticipate potential issues (i.e., software defects) given their limited programming experience and what are the mistake that students might have when anticipating issues for code reviews.





In this paper, we aim to investigate whether students can anticipate potential issues before conducting a code review.
Hence, we create an assignment to ask students to develop a code review checklist based on a given requirements specification.
Then, we examine the student's checklists with respect to two research questions: (1) \rqOne{} and (2) \rqTwo{}.
To answer the research questions, we conduct a manual categorization of the questions in the students' checklist into the common defects.
We also perform an open coding to identify the common mistakes that the students make when developing a code review checklist. 
We perform a study with 394 students over the 2018 and 2019 cohorts at \monash{} across two campuses (i.e., Australia and Malaysia).
The results of this study will help educators improve the teaching strategies to stimulate students in developing their analytical skills for code reviews.

The structure of the paper is as follow.
Section \ref{sec:background} provides  a background of the code review practice and the importance of code review.
Section \ref{sec:relatedwork} discusses the related work.
Section \ref{sec:methodology} describes our research methodology.
Section \ref{sec:results} presents the results.
Section \ref{sec:discussion} discusses a broader implication of the results and the limitations of our study.
Finally, Section \ref{sec:conclusion}  draws a conclusion.

\section{Background}
\label{sec:background}
In this section, we provide a background of the code review practice and software defects that are commonly found during code reviews.

\subsection{Code Review}
Code review (or software inspection) is one of the most important practices for software quality assurance which is the core foundation of the Software Engineering discipline. 
Based on Linus's law \textit{``Given enough eyeballs, all defects are shallow''}~\cite{raymond1999cathedral}, the main goal of code review is to manually examine source code to identify software defects, i.e., an error, flaw or fault that causes a software system to behave incorrectly or produce an unexpected result.
Several studies have shown that code review provides substantial benefits to improve the quality of a software system~\cite{thongtanunam2015investigating,Shull2002,fagan2001advances}. 
For example, Shull~\ea~\cite{Shull2002} argued that code reviews can uncover more than half of software defects which is more than the number of defects found from software testing. 
Fagan~\cite{fagan2001advances} also found that code review at IBM can uncover 90\% of software defects over the lifecycle of a product.
A more detail of software defects is described in Section \ref{sec:defect_type}.
In addition, prior studies also showed that code review provides additional benefits to software engineering practices, such as improving team cohesion and communication~\cite{sripada2015support}, easing the maintainability of the software system~\cite{mantyla2008types,Morales2015}, and knowledge sharing~\cite{thongtanunam2016revisiting}.

\begin{figure}[t]
\centering
\includegraphics[width=.85\columnwidth]{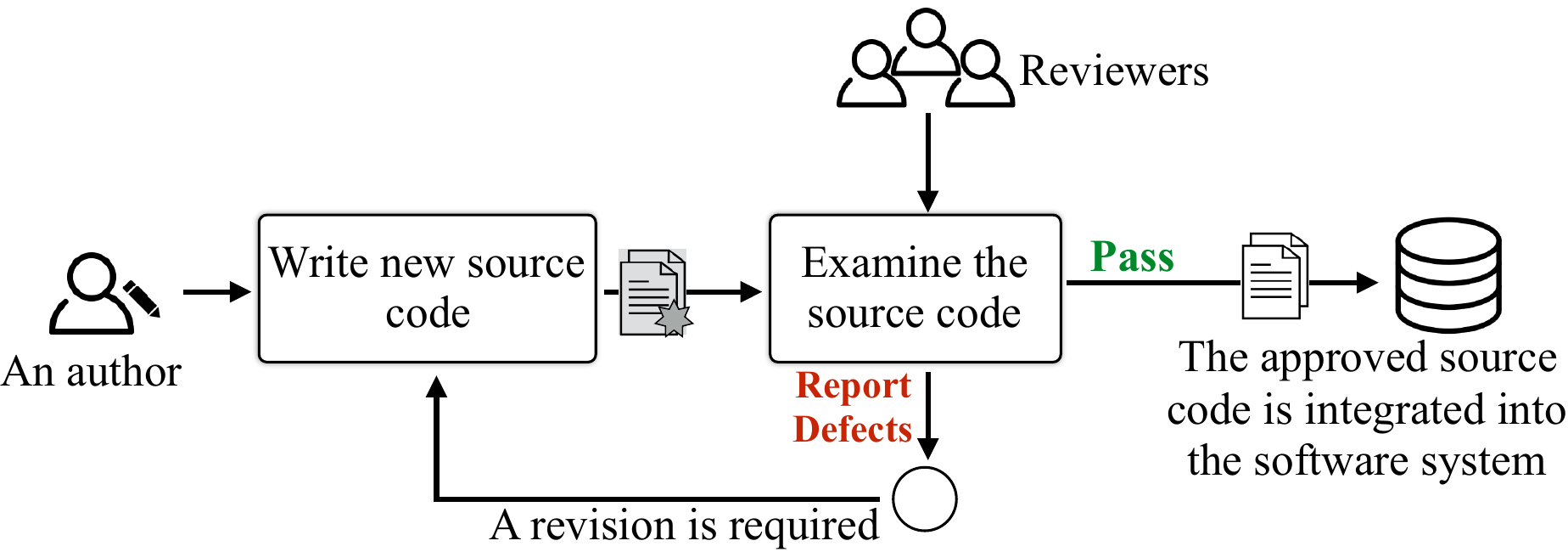}
\caption{An overview of the code review process.}
\label{fig:code_review_process}
\end{figure}

Figure \ref{fig:code_review_process} shows an overview of the code review process.
In general, when an author (i.e., a software developer) writes new pieces of source code for a software system, reviewers (i.e., software developers other than the author) will carefully examine the source code and identify potential defects or problems that new source code might unintentionally introduce to the existing codebase.
If defects are indeed found, the author will revise the source code to address the defects or potential problems reported by the reviewers.
Once the reviewers agree that the new source code is of sufficiently high quality and free of observable defects, the new source code will be integrated into the software system.

To effectively identify defects, code review requires the analytical and reading skills of reviewers~\cite{Aurum2002,sripada2015support}.
Adhoc and checklist-based code reviews are the common methods that reviewers use when examining source code.
Below is a brief description of the ad-hoc and checklist-based code review techniques.

\begin{itemize}
\item \textbf{Ad-hoc Code Review} is a lightweight method that is widely used in modern software organisations. The code review of this method is based on a general viewpoint of reviewers who use their personal experience and expertise to identify defects.
The strength of this method is that it gives the reviewers freedom to read
the code and identify various defects~\cite{laitenberger2000encompassing}.
On the other hand, the weakness in this method is that code review can be ineffective if it is performed by inexperience reviewers.

\item \textbf{Checklist-based Code Review} is a more systematic method of code review.
The method is proposed by Fagan~\cite{Fagan1976}, where a reviewer prepares a checklist (i.e., a list of \checklistItem{s} or predefined issues or defects that need to be checked) beforehand.
Then, the checklist is used to guide the reviewer to identify defects when performing code review.
A strength of this method is that a checklist will help reviewers to concentrate their focus on important issues or defects~\cite{Aurum2002}.
A weakness of this method is that inexperienced reviewers may overlook the defects that are not directed by the checklist.
\end{itemize}

\input{sections/defect_type.tex}

\subsection{Software Defects in Code Reviews}
\label{sec:defect_type}
Prior work attempted to categorise software defects that are commonly found in source code during code reviews~\cite{mantyla2008types}.
They argued that defect classification can be useful when creating company coding standards and code review checklists.
Moreover, recent studies also used the taxonomy proposed by Mantyla~\ea~\cite{mantyla2008types} to evaluate the recent practices of code reviews in open source projects~\cite{Beller2014,thongtanunam2015investigating}.
Table \ref{Table:taxonomy} summarises software defects in code review, which are briefly described as follow.
\textbf{Documentation defects} refer to the issues on the textual information that eases the understanding of source code (e.g., element naming and comments) or that is used to document the run-time problems (e.g., exception messages). The concern also includes programming language matters related to documentation (e.g., the scope of a variable and method).
\textbf{Visual representation defects} refer to the concern on the readability of source code. 
The concern mainly focuses on how should the code elements be organised (e.g., removing unnecessary blank spaces) to ease developers when reading and comprehending the source code.
\textbf{Structure defects} refer to the concern on the maintainability and evolability of the inspected software. Unlike the documentation and visual representation, structure defect types also includes compilation and run-time issues.
\textbf{Resource defects} refer to the defect of incorrect manipulation of data and other resources. The concern also includes memory allocation and release for variables.
\textbf{Check defects} refer to the defect types related to incorrect or missing validation of variables or values. This concern also includes the validation check of user inputs.
\textbf{Interface defects} refer to the concern on the use of APIs.
\textbf{Logic defects} refer to the concern on the correctness of logic in the system.
\textbf{Large defects} refers to defect types on incomplete features and user interfaces.

\section{Related Work}
\label{sec:relatedwork}
\subsection{Teaching Code Reviews for Software Engineering Education}
In software industry, code review has become a mandatory step in the Software Engineering (SE) workflow of many software organisations, e.g., Google~\cite{sadowski2018modern}, Microsoft~\cite{Rigby}. 
Hence, it is imperative for SE students to be equipped with code review skills prior to their employment.  
Code review activities have been integrated into the software engineering and computer science syllabus in many universities~\cite{sripada2015support,wang2008process,hundhausen2009integrating}.
For example, Wang~\ea~\cite{wang2008process} and Hundhausen~\ea~\cite{hundhausen2009integrating} integrated code review activities into the Computer Science subject called ``pedagogical code review'' where students assess and provide feedback for a software program of other students. 
Portugal~\ea~\cite{portugal2016facing} showed that code inspection can be used to validate the functional and non-functional requirements elicited from stakeholders when introduced at the university-level. 

From a pedagogy's point-of-view, Hilburn~\ea~\cite{hilburn2011read} suggested that code review is an active learning technique in software engineering education because it is a team activity; it requires technical knowledge to participate; and it involves measurement and use of data to analyse and draw conclusions. Kemerer and Paulk \cite{kemerer2009impact} found that when code review activities are included in the learning activities, the quality of student submissions tend to be improved.
Hundhausen~\ea~\cite{hundhausen2009integrating} also reported that practicing code review helps students improve the quality of their code and stimulates meaningful discussion and critiques.  

\subsubsection{Challenges in Teaching Code Review}

Teaching effective code review is challenging from an educational perspective since code review requires a programming experience, code reading skills, and analytical skills. 
Prior works mainly focuses on investigating the code reading techniques that help students effectively find defects~\cite{mcmeekin2009evaluating,rong2012effect,sripada2015support}. 
For example, McMeekin~\ea~\cite{mcmeekin2009evaluating} found that ad-hoc code review requires higher cognitive load and it is only suitable for students with a higher education level or have experience in programming.
On the other hand, checklist-based code review is suitable for less experienced students as it can be used to guide code reading of the students.
Rong~\ea~\cite{rong2012effect} found that a checklist helps students in terms of a guidance for reading code, i.e. allow students to focus on the important tasks at hand, within a stipulated amount of time.
Several code review checklists are developed in order to support code reading during the code reviews exercise~\cite{dunsmore2003practical,Humphrey:1995:DSE:526070}.

In addition to the code reading skills, students should learn how to develop a code review checklist.
In other words, providing a checklist to students may hinder the students from learning to analyse and anticipate potential issues or defects that might arise in code review. 
Moreover, using a generic checklist may lead students to overlook the context-specific problems~\cite{mcmeekin2009evaluating}.
Hence, developing a code review checklist should be a learning activity that helps students develop necessary analytical skills to anticipate problems in the context of software development.
Furthermore, the industrial practices recommend that reviewers should create and use their own checklists for code reviews~\cite{MacLeod2018}.
Moreover, it is generally acknowledged that the quality of the checklist will predetermine the quality of the code review process \cite{dunsmore2003development}.
However, Rong~\ea~\cite{rong2012effect} argued that developing a high quality checklist is a difficult task for students because students typically have limited domain knowledge (e.g., programming, design, or testing skills).
Yet, it is still unclear whether students can develop a code review checklist (i.e., a list of questions that guide reviewers to look for defects) given their limited programming experience.
The findings of this work are of importance to help educators to provide early guidance to students if mistakes or misconceptions were found in the preliminary phase of code review process.

\section{Research Methodology}
\label{sec:methodology}
In this section, we describe the goal of this work, the research questions, and the design of our study.


\subsection{Goal \& Research Questions}

The goal of this work is to investigate whether students can develop a code review checklist (i.e., a list of \checklistItem{s} that guide reviewers to look for defects). 
To achieve this goal, we formulate the following research questions:

\begin{itemize}
    \item \textbf{RQ1: \rqOne{}}  \\
    \underline{Motivation:} This RQ aims to assess whether students can apply analytical skills to create checklists as a guideline to identify software defects during code reviews. The findings will shed some light on the feasibility of this learning activity to educate students to perform effective code reviews.\\
    \underline{Measurements:} We examine (1) the number of \checklistItem{s} in the students' checklists that can be used to guide reviewers to uncover software defects and (2) the types of software defects that can be identified. 
    
    \item \textbf{RQ2: \rqTwo{}}  \\
    \underline{Motivation:} This RQ aims to investigate common mistake among students when developing a code review checklist. The findings could help educators to be aware of and prepare themselves to address the potential mistakes or misunderstanding that students might have when performing code reviews.\\
    \underline{Measurements:} We examine (1) the number of \checklistItem{s} in the students' checklists that have poor quality and those questions that should not be used to identify software defects and (2) the characteristics of the common mistakes in the students checklists.
\end{itemize}

    

\subsection{Designing a Checklist-based Code Review Assignment}

To conduct the study, we designed an assignment for developing a code review checklist based on the requirements specification of a given software system.
Below, we describe the code review assignment and the teaching and workshop activities that help students to achieve the objective of the code review assignment.

\subsubsection{The Code Review Assignment}

\begin{figure}[t]
\centering
\includegraphics[width=\columnwidth]{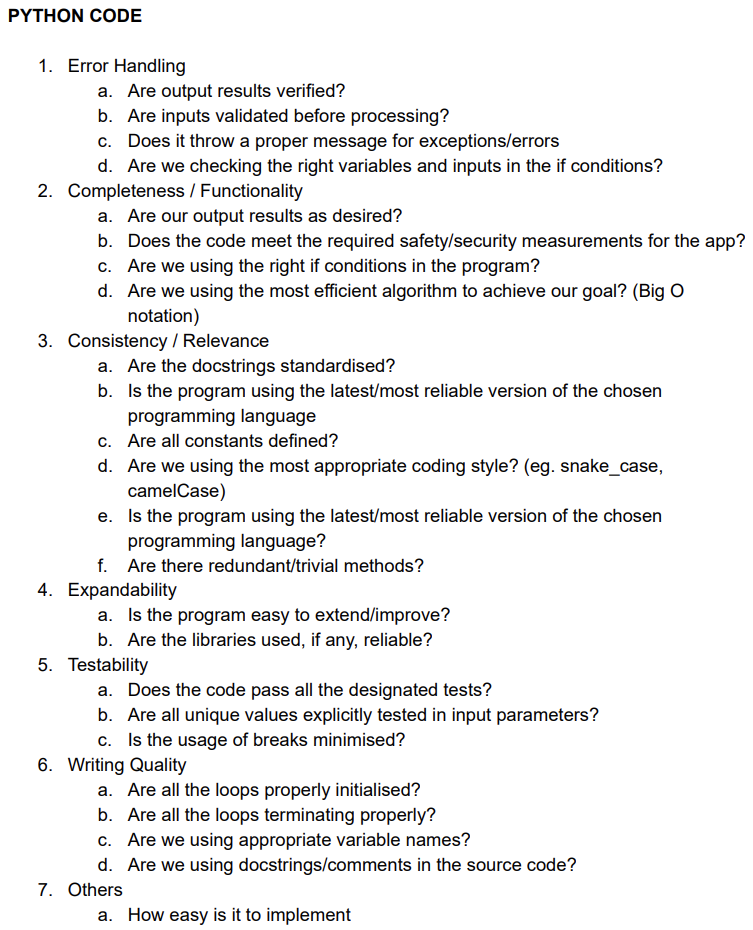}
\caption{\normalsize An example of a code review checklist submitted by students.
\label{fig:studentchecklist}}
\end{figure}

The objective of the code review assignment is to assess students' understanding of checklist-based code review activities (i.e., how do students develop code review checklists).
Generally speaking, the students are required to conduct a formal code inspection meeting in a group of 4-5 students in order to mimic formal code review practices on a set of given software artifacts.
The structure of the assignment is as follows:

\begin{enumerate}
     \item Week 2: Students are required to create and submit code review checklists (individually) for a given software artifact. Noted that at this stage, the students only have access to the requirements specification, without the actual artifact that needs to be reviewed. The requirements specification and relevant software artifacts are available online \cite{chong_chun_yong_2021_4431059}. We do not set a limit on the number of checklist questions that the students can prepare. This will allow the students to design their checklists based on the requirements (i.e. programming languages, expected behaviour of the system, etc.). An example of student's submissions is shown on Figure \ref{fig:studentchecklist}.
     \item Week 3: After submitting their checklists, they will be given an actual artifacts to be reviewed. Students need to individually identify potential quality issues (i.e., defects) by applying their own checklists on the provided software artifact. 
     Each student will then prepare a list of issues that they identified, based on their constructed checklist.
     \item Week 4: Students are required to conduct a formal inspection with their group members in one of the workshop sessions to discuss and report software defects that they discovered. Students will take turn and discuss the issues that they have discovered, and in return justify if their constructed checklists are useful in identifying potential software defects.
 \end{enumerate}


\subsubsection{Teaching Activities}
During the lecture, we highlight that due to the costliness of manual code review, reviewers are required to focus on the most important tasks at hand. Thus, code review checklist plays a very important role to facilitate the review process.

The subject is structured around semi-flipped learning environment, where students are expected to go through pre-class reading notes before attending the weekly lectures. The pre-class reading notes provide detail discussion on the week's topic, with an estimated reading time of 30-45 minutes. The weekly lectures (60 minutes per week, 12 weeks in total) will emphasis on the core deliverable of the week and less time are spent on discussing the definition and terminologies since students are expected to go through the reading notes before attending the lectures. Apart from that, there are 2-hour weekly workshops where the students will conduct hands-on exercise on the weekly topics.

In Week 2 of the workshop, students will start working on the Code Review Assignment. They will be doing individual research and preparation to come out with their own sets of checklists that are relevant to the artifact to be reviewed, which in this case is a piece of code written in Python. In Week 4 of the workshop, students will bring along their constructed checklist items, the list of defects that they have discovered using their own checklist, and conduct the face-to-face code review meeting together with their teammates.

\subsection{Data Analysis}

We manually analyze the quality of students' checklists.
To do so, we first manually classify the \checklistItem{s} in the students' checklists based on the types of functional defects shown in Table \ref{Table:taxonomy} that we obtained from Week 2.
For the questions that cannot be classified into the functional defects, we perform an open coding to define non-functional defects and common mistakes. 
We noted that the study protocol has been rigorously reviewed and approved by the \monash~Human Research Ethics Committee (MUHREC Project ID: 21958).

Below, we describe the details of the two phases of our data analysis and the context of the studied cohorts.

\subsubsection{Data Analysis-1: Manual Classification of Defect Types}
A good review checklist should have \checklistItem{s} that guide reviewers to look for specific defects during code review activities.
Hence, we analyze the type of defects that the \checklistItem{s} in the students' checklist aim to identify.
To evaluate the students' checklists, we use the taxonomy of functional defects found during code reviews of Mantyla~\ea~\cite{mantyla2008types} (see Table \ref{Table:taxonomy}).
The classification is performed by a Research Associate (RA) with solid  experience in developing checklists and performing code reviews and teaching our software quality and testing subject in the past 3 years.
To ensure the validity of the results, the first author examined the classification results once the first half of the \checklistItem{s} of the students' checklists are classified. 
After the validation, the RA revisited the classified \checklistItem{s} to ensure a consistent understanding, then continue classifying the remaining \checklistItem{s} in the checklists. 
Finally, the first author examined all the results.
The checklist \checklistItem{s} can be classified as undefined if those \checklistItem{s} are not relevant to any predefined defects.
These undefined \checklistItem{s} will be further analyzed in the next step.

\subsubsection{Data Analysis-2: Open Coding for the Undefined Category}
The defect types proposed by Mantyla~\ea~\cite{mantyla2008types} are mainly for functional defects. However, other non-functional defects can be included in the checklists~\cite{portugal2016facing}. 
Hence, we manually analyze the \checklistItem{s} in the undefined category of the Data Analysis 1 to see whether they can be used to guide reviewers to look for potential quality issues. 
We use an open coding approach similar to prior studies~\cite{Rigby2011a,hirao2019review}.
The coding was performed by all the authors of this paper during online meeting sessions.
All the coders have experience with code review research and and have strong experience in teaching software engineering subjects. 
The coders identify the key concerns of each question in the checklist. 
The key concerns are then noted down when all the coders reach a consensus.
In addition to identifying key concerns, the coders also determine whether or not the checklist \checklistItem{s} are appropriate, i.e., cannot be used to guide reviewers to identify software defects. 
If the checklist question is not appropriate, the coder identify the type as a mistake.
All the coders manually identify the key concerns and mistakes for each of the checklist \checklistItem{s} in the undefined category until reaching a \textit{saturation} state, i.e., no new defect types or mistakes are discovered after coding 50 consecutive checklist \checklistItem{s}.
Once the list of newly discovered defect types reaches the saturation state, the first author continues to identify the key defect types and the mistakes of the remaining checklist \checklistItem{s}.
Finally, the second author revisited these \checklistItem{s} to validate the results.
If a disagreement occurs, all the coders discuss until a consensus is reached.

\subsection{Studied Cohorts and Context}
The participated students are Software Engineering students who enrolled the Software Quality and Testing subject at \monash~in 2018 and 2019.
The Software Quality and Testing subject is offered as a second-year core subject at an undergraduate level as part of a 4-year Bachelor degree in Software Engineering.
The program's curricula and learning outcomes are aligned with the recommendation from national and international bodies such as ACM and IEEE.
The subject is offering for two campuses at different countries (i.e., the main campus is in \australia~and the branch campus is in \malaysia).
Students of both campuses will undertake the same subject structure, assignments, exams, teaching, and learning materials, ensuring the subject is delivered at the same quality for both campuses.
For the main campus, there are 128 enrolled students for 2018 and 159 enrolled students for 2019.
For the branch campus, there are 40 enrolled students for 2018 and 67 enrolled students for 2019. 
Other characteristics of the participated students are described below.

\emph{English Language Proficiency:}
\monash~is an English-speaking institution and all participants have basic competence in English to take part in the study. 

\emph{Programming Experience:} 
The prerequisite of this subject is either algorithms and programming fundamentals in Python or programming fundamentals in Java, indicating that students have enough programming experience to perform code review.

\emph{Code Review Knowledge:} Before working on the assignment, the students were given a one-hour lecture of the code review process, including the importance and main goals of conducting code reviews. An example of generic code review checklists~\cite{dunsmore2003practical,Humphrey:1995:DSE:526070} is also presented during the lecture hour.

\begin{figure*}[t]
\centering
\includegraphics[width=\textwidth]{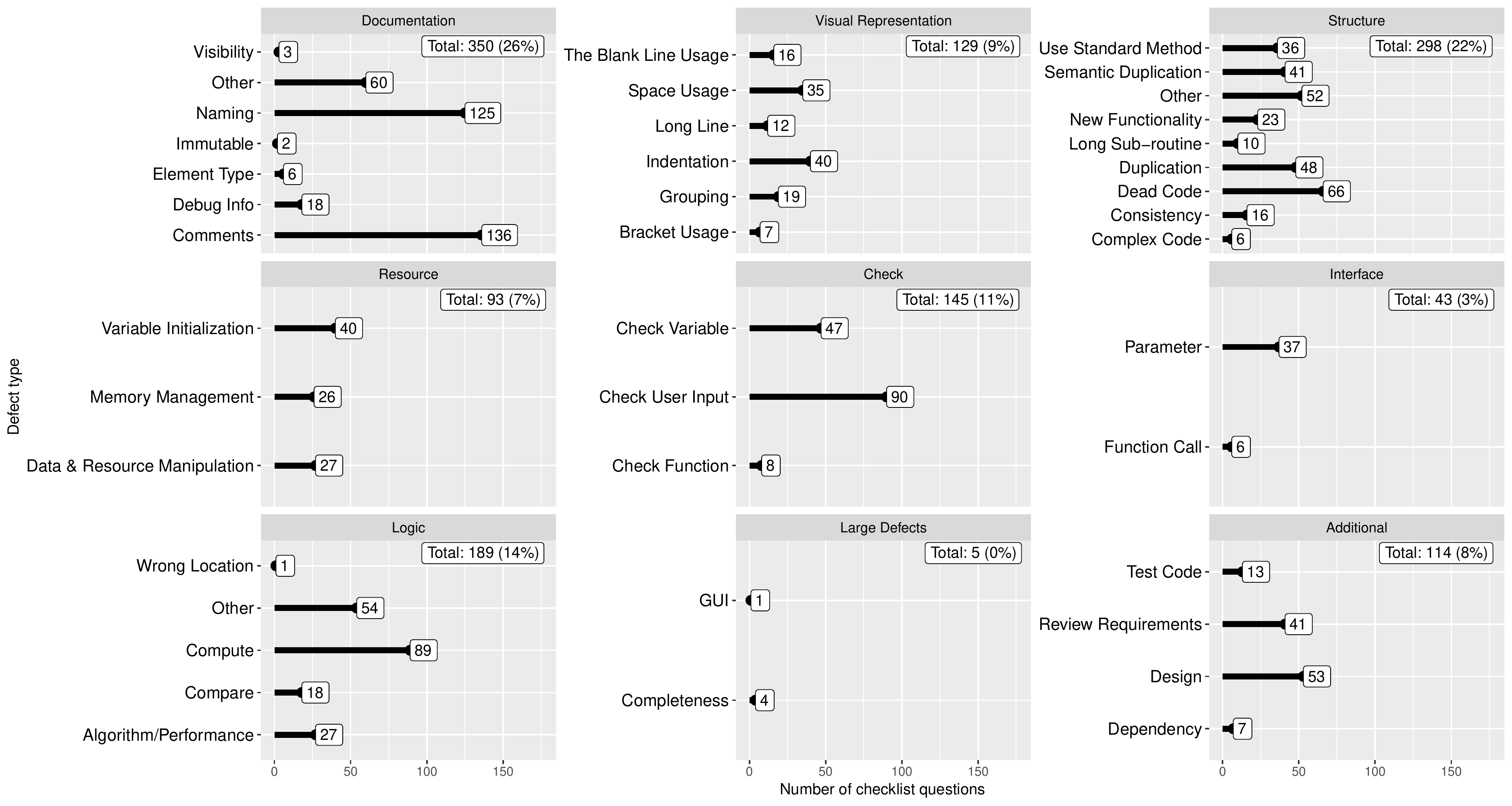}
\caption{\normalsize The number of questions in the checklists with respect to the Mantyla's \cite{mantyla2008types} taxonomy of defect types and the additional categories.
\label{fig:suitable}}
\end{figure*}

\section{Analysis Results}
\label{sec:results}

We received a total of 1,791 \checklistItem{s} from 394 students across the two studied cohorts in 2018 and 2019.
Below, we present the analysis results with regarding to our research questions.

\subsection*{RQ1: \rqOne}
We find that 76\% of the checklist \checklistItem{s} can be used to identify defects.
More specifically, 70\% of the checklist \checklistItem{s} that are produced by students are clear and can be used to guide reviewers to identify the common defects listed in Table \ref{Table:taxonomy}.
In addition, we find that another 6\% of the checklist \checklistItem{s} aim to identify non-functional defects.
A relatively large proportion of appropriate checklist \checklistItem{s} suggests that students can anticipate potential problems and produce a relatively good quality of code review checklist.

Figure~\ref{fig:suitable} shows a distribution of the checklist \checklistItem{s} across the defect types.
Note that the proportion shown in the figure is the proportion of appropriate checklist \checklistItem{s}.
The result shows that documentation (26\%) and structure (22\%) defects are the majority defect types that the \checklistItem{s} of students checklist aim to identify.
The ratio between functional defects (i.e., resource, check, interface, logic, and large defects) and other defects in the student checklists is 35:65, which is similar to the ratio of actual defects found during code review conducted by practitioners (25:75) \cite{mantyla2008types,Beller2014}.
This result suggests that student checklists should guide reviewers to uncover defects with a similar ratio of defects as the literature.

In addition to the common defects listed in Table \ref{Table:taxonomy}, we find that student checklists also include questions that can identify design, requirement, test code, and dependency defects. 
Based on the recent studies, these defects can be uncovered in the code review process~\cite{he2008using,portugal2016facing,spadini2018testing}.
These additional types of defects account for 8\% of the appropriate \checklistItem{s} in the students' checklists (see Figure~\ref{fig:suitable}).
\emph{Design defects} refer to the defects about the consistency between design and code artifacts.
\emph{Requirement defects} refer to the defects about the clarity and completeness of code artifacts regarding the requirements specification.
\emph{Test code defects} refer to the defects about the quality of the test cases such as test coverage, completeness of the test case documentations (test conditions, test environment, test inputs, expected outcomes, etc.), traceability to requirements, and reusability of test code. 
\emph{Dependency defects} refers to defects about the usage of external libraries, modules, or APIs as dependencies. 
This result suggests that in addition to the common defects, students also can anticipate defects in other aspects.

\begin{table}[t]
\caption{The distribution of appropriate \checklistItem{s} across defect types.}
\label{Table:cohorts}
\centering
\begin{tabular}{l|rr} 
\hline
\textbf{Defect Type} & \textbf{2018} & \textbf{2019} \\
\hline
Documentation  &        29\%  & 23\% \\ 
Visual Representation & 6\% & 12\% \\
Structure     &        24\%  &  20\% \\
Resource      &         5\% &  8\% \\
Check         &        11\%  & 11\% \\
Interface     &         1\% & 5\% \\
Logic         &        15\%  & 13\%  \\
Large Defects  &        1\% & 0\% \\
Additional     &        9\%  & 8\% \\
\hline
\end{tabular}
\end{table}

Since we analyze the students' checklists from the 2018 and 2019 cohorts, it is possible that the results may be different between the two cohorts.
Table \ref{Table:cohorts} shows a proportion of appropriate checklist \checklistItem{s} in each defect type for each cohort.
We perform the Pearson's chi-squared test ($\alpha = 0.05$), i.e., a statistical test for independence of categorical variables, to examine whether the distributions of the checklist \checklistItem{s} for the 2018 and 2019 cohorts are statistically different.
The test result shows that the distributions of the checklist \checklistItem{s} for the two cohorts are not statistically different with $p$-value of 0.6306. 
This suggests that the checklists produced by the students in these two cohorts are not statistically different.

\begin{figure}[t]
\centering
\includegraphics[width=\columnwidth]{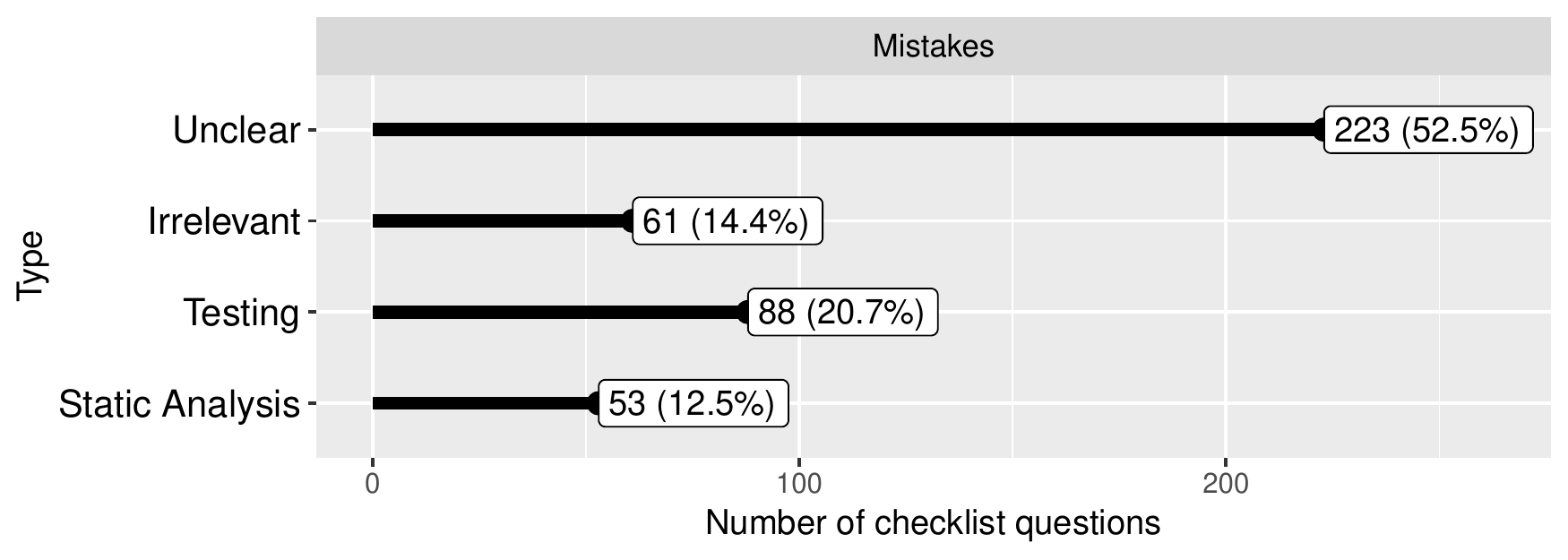}
\caption{\normalsize The number of inappropriate \checklistItem{s} in the students' checklists.
\label{fig:misconception}}
\end{figure}

\subsection*{RQ2: \rqTwo}
From 425 inappropriate checklist \checklistItem{s} (26\% of all checklist \checklistItem{s}), we can identify four common mistakes, i.e, (1) unclear, (2) irrelevant, (3) testing, (4) static analysis.
Figure~\ref{fig:misconception} shows that \textit{unclear} is the most common mistake (52\%) and the other three mistakes account for a similar proportion (12.5-20.7\%). 
Below, we provide a description, an example and our rationale for each mistake type.

\textbf{Unclear checklist \checklistItem{s}} refer to the checklist \checklistItem{s} that are too generic or ambiguous to guide reviewers to find a defect. 
Brykczynski \cite{brykczynski1999survey} also discussed that a general checklist is inappropriate for code review because it will hinder the reviewer to pay attention on a particular area of the code, which could result in discovering insignificant defects. 
For example, ``\textit{Is the code maintainable?}'' is considered as unclear checklist \checklistItem{s}.
This is because maintainability is a broad aspect and cannot be directly quantified. 
Hence, such kind of questions are impractical for reviewers to identify defects.
To assess the code maintainability, the checklist \checklistItem{s} should focus on the defects in the documentation, structure, visual representation aspects listed in Table\ref{Table:taxonomy}.


\textbf{Irrelevant checklist \checklistItem{s}} refer to the checklist \checklistItem{s} that are irrelevant from the context, e.g., the given requirements specification or programming languages.
For example, ``\textit{Are all shared variables minimised and handling concurrency of multiple threads and is not leading to a potential deadlock situation?}'' is considered irrelevant to the given context.
This is because the the use of multiple threads was not described in the requirements specification and the assignment details.
We suspect that such irrelevant \checklistItem{s} were derived from a generic checklist that is available online. 

\textbf{Testing-dependent checklist \checklistItem{s}} refer to the checklist \checklistItem{s} that are dependent to the testing results.
For example,``\emph{Does the code pass unit, integration, and system tests?}'' is considered as an inappropriate question.
This is because code review should focus on defects based on the source code. 
Code review should be a complimentary software assurance activity that can discover defects other than the defects detected by testing.
Hence, code review should not be dependent to the testing results.

\textbf{Static Analysis-dependent checklist \checklistItem{s}} refer to the checklist \checklistItem{s} that are dependent to the results of the static analysis tool that check for code styling or syntax, for example, ``\textit{Does Pylint give a reasonable score (at least 7/10)?}''.
This kind of checklist \checklistItem{s} are considered as inappropriate because code review should focus on more in-depth defects.
Moreover, such errors should be removed by the developers prior to the code review process to minimize the code review effort.

\section{Lessons Learned}
\label{sec:discussion}
Our analysis results show that given a basic knowledge of algorithms and programming, 76\% of the checklist \checklistItem{s} are appropriate which can be used to guide reviewers to find defects.
The results suggest that students tend to be able to anticipate potential defects and create a relatively good code review checklist.
Nevertheless, students might make some mistakes when developing a code review checklist.
The first two common mistakes (i.e., unclear and irrelevant questions) are in part due to the misconception about the quality specification of the code review checklist.
Thus, we recommend educators cope this issue in the future by making a clear specification of the checklist.
The other two common mistakes (i.e., testing-dependent and static-analysis-dependent) are in part due to the misconception about the goal of code review process.
Thus, we recommend educators emphasize the key role of code review in software engineering process and a clear distinction of software quality assurance activities.

In addition, developing a code review checklist should be included as one of the learning activities for code review.
This can be considered as a scaffolding activity. 
Students can demonstrate their understanding about code reviews through the development of a code checklist.
Then, educators can give a guidance or correct the misconception that students might have before reading code and identifying defects during a code review.  
For example, we observed that the checklist \checklistItem{s} are more focused on documentation, while the defects found in the industrial reviews are structure and large defects~\cite{mantyla2008types}. 
Then, before conducting a code review, educators should support students by emphasizing more on the structural defects in order to stay in line with the industrial practices.

\textbf{Limitation.}
Our findings shed the light on the possibility of including code review checklist as an additional learning activity for code review assignment.
However, there are some limitations with the research methodology and the consideration to apply the research results in general.
The first limitation is the diversity of participants.
Although the data is collected from a large group of participants (394 students) from two years of cohorts (2018 and 2019) and two campuses, all the participants are from one university.
The conclusion of this study may not be generalized to the students of other universities.
Nevertheless, our prerequisite is common, i.e., students have passed algorithms and programming fundamentals. 
Hence, it is feasible to revisit this work in a different university.
The second limitation is the programming background of students.
The participants are second-year undergraduate students with an algorithm and programming background.
Hence, the conclusion may only be generalized to undergraduate students with limited programming background.
Students with more software engineering experience and stronger programming background may generate a different code review checklist.

\section{Conclusion}
\label{sec:conclusion}
In this work, we investigate whether students can anticipate potential issues when performing code review.
To do so, we examine the code review checklists (a list of \checklistItem{s} that guide reviewers to look for defects) that are developed by students.
From the 1,791 checklist \checklistItem{s} of 394 students across the two cohorts at \monash, we find that 76\% of the questions are appropriate to be used to guide reviewers during code review.
More specifically, 70\% of the checklist \checklistItem{s} can be used to guide reviewers to identify defects that are commonly found during the code review.
We also identified four common mistakes of the checklist \checklistItem{s} that educators should be aware of when asking students to develop a code review checklist.
Our results shed the light on an additional learning activity of code reviews, i.e., developing a code review checklist, which should stimulate students in developing their analytical skills for code reviews.

\section*{Acknowledgement}
C. Tantithamthavorn was partially supported by the Australian Research Council's Discovery Early Career Researcher Award (DECRA) funding scheme (DE200100941).
P. Thongtanunam was partially supported by the Australian Research Council's Discovery Early Career Researcher Award (DECRA) funding scheme (DE210101091).

\bibliographystyle{IEEEtranS}
\bibliography{filteredref} 

\end{document}

%% file: sections/defect_type.tex
\begin{table*}[t]
\caption{A Taxonomy of Defects in Code Review~\cite{mantyla2008types}}
\label{Table:taxonomy}
\centering
\resizebox{\linewidth}{!}{
\begin{tabular}{ll} 
\hline
Type & Description \\
\hline
\multicolumn{2}{l}{\textbf{Documentation Defects}}\\
\hline
Naming & Defects on the software element names. \\
Comment & Defects on a code comment. \\
Debug Info & Defects on the information that will be used for debugging. \\
Element Type & Defects on the type of variables. \\
Immutable & Defect when not declaring variable to be immutable when it should have been.\\
Visibility & Defects on the visbility of variables. \\
Other & Other defects relating to textual information.\\
\hline
\multicolumn{2}{l}{\textbf{Visual Representation Defects}}\\  
\hline
Bracket Usage & Defects on incorrect or missing necessary brackets. \\
Indentation & Defects on incorrect indentation. \\
Blank Line Usage & Defects on the use of incorrect or unnecessary blank lines. \\
Long Line & Defects on a long code statement which cause the readability. \\
Space Usage & Defects on unnecessary blank spaces. \\
Grouping & Defects on the cohesion of code elements. \\
\hline
\multicolumn{2}{l}{\textbf{Structure Defects}}\\  
\hline
Long Sub-routine &  Defects on a lengthy function or procedure. \\
Dead Code & Defects on code statements that do not serve any meaningful purpose. \\
Duplication & Defects on duplicate code statements. \\
Semantic Duplication & Defects on code redundancy. \\
Complex Code & Defects on source code that are difficult to comprehend. \\
Consistency & Defects on the ways of implementation should be in a similar fashion.\\
New Functionality & Defects on a piece of source code that does not have a single purpose. \\
Use Standard Method & Defects whether the standardized approach should be used. \\
Other & Other defects related to refactoring, organizing, and improving source code.\\
\hline
\multicolumn{2}{l}{\textbf{Resource Defects}}\\  
\hline
Variable Initialization & Defects on the variables that are uninitialized or incorrectly initialized. \\
Memory Management & Defects on how program use and manage the memory. \\
Data \& Resource Manipulation & Other defects on the resource management. \\
\hline
\multicolumn{2}{l}{\textbf{Check Defects}}\\  
\hline
Check Function & Defects on the returned value of a function. \\
Check Variable & Defects about whether the program validate the variable values.\\
Check User Input & Defects on the validity of user input. \\
\hline
\multicolumn{2}{l}{\textbf{Interface Defects}}\\  
\hline
Parameter & Defects on incorrect or missing parameters when calling another function or library. \\
Function Call & Defects on incorrect interaction between a function and another part of the system or class library.\\
\hline
\multicolumn{2}{l}{\textbf{Logic Defects}}\\  
\hline
Compare & Defects on logic for a comparison. \\
Compute & Defects on incorrect logic when the system runs. \\
Wrong Location & Defects on the wrong location of the operation although it is correct. \\
Performance & Defects on the efficiency of the algorithm.\\
Other & Other defects related to the correctness or existence of logic.\\
\hline
\multicolumn{2}{l}{\textbf{Large Defects}}\\  
\hline
GUI & Defects about the completeness and correctness of user interfaces. \\
Completeness & Defects about whether the implementation is complete.\\
\hline
\end{tabular}
}
\end{table*}